%% file: main.tex
\documentclass[journal]{IEEEtran}

\usepackage[T1]{fontenc}
\usepackage{amsmath,amssymb,bm}
\usepackage{graphicx}
\usepackage{cite}

\IEEEoverridecommandlockouts

\AtBeginDocument{%
  \setlength{\abovedisplayskip}{3pt plus 1pt minus 1pt}%
  \setlength{\belowdisplayskip}{3pt plus 1pt minus 1pt}%
  \setlength{\abovedisplayshortskip}{2pt plus 1pt}%
  \setlength{\belowdisplayshortskip}{2pt plus 1pt}%
}

\newcommand{\R}{\mathbb{R}}
\newcommand{\smin}{\sigma_{\min}}
\newcommand{\pmap}{\Pi}

\begin{document}

\title{Loadability Limits Under Periodic Load Forcing}

\author{Maitraya~Avadhut~Desai,~\IEEEmembership{Graduate~Student~Member,~IEEE,}
    Gustavo~Valverde,~\IEEEmembership{Senior~Member,~IEEE,}
    and Gabriela~Hug,~\IEEEmembership{Senior~Member,~IEEE}%
    \vspace{-1.15cm}
    \thanks{This research was supported by the Swiss National Science
        Foundation under NCCR Automation, grant agreement
        51NF40\_180545.}%
    \thanks{M.\,A.\,Desai, G.\,Valverde and G.\,Hug are with the Power
        Systems Laboratory at ETH Zurich, 8092 Zurich, Switzerland, emails:
        \{desai, valverde, hug\}@eeh.ee.ethz.ch. \textit{(Corresponding
            Author: G.\,Valverde)}}%
}

\maketitle
\IEEEpeerreviewmaketitle

\begin{abstract}
    \looseness=-1 The static loadability limit, defined as the demand at which the
    equilibrium equations lose their solution, is the standard basis for interconnection
    screening of large new loads. This letter shows that when part of
    the demand varies periodically, as for data-center loads, the
    steady state is a forced periodic orbit whose loadability limit
    differs from the static one. The
    classical optimization argument is extended directly from
    equilibria to fixed points of the period map. At the limit, the
    monodromy matrix acquires a Floquet multiplier at $+1$, so the
    generic instability is a cyclic fold of the orbit rather than a
    saddle-node of equilibria. The limit is therefore a function of
    the forcing frequency, which no static computation can capture.
    Additionally, with the network Jacobian turning singular along the
    cycle, singularity-induced instability is extended to orbits. On a
    four-bus test system, a static margin of 2.5~p.u.\ shrinks to
    0.53~p.u.\ near the swing frequency, the instability mechanism switches
    from voltage collapse to a rotor-angle fold, and cold starts fail
    at amplitudes where the orbit still exists. A static analysis
    reproduces none of these effects.
\end{abstract}


\vspace{-3mm}
\input{sections/introduction}
\input{sections/theory}
\input{sections/casestudy}
\input{sections/conclusion}

\bibliographystyle{IEEEtran}
\bibliography{refs}

\end{document}

%% file: sections/introduction.tex
\section{Introduction}

\looseness=-1 \IEEEPARstart{D}{ata} centers are among the largest new loads
seeking interconnection to the power grid. Their synchronized computing workloads
draw power that pulsates at well-defined frequencies, with swings of a
large fraction of the facility rating~\cite{Chalamala2025}.
Interconnection studies, however, screen loads against the
\emph{static} loadability limit, the maximum demand for which the network
equations retain a solution.

\looseness=-1 Static loadability limits generically coincide with
saddle-node bifurcations, where the Jacobian of the full equilibrium equations is singular,
sensitivities diverge, and the admissible demands are bounded by a bifurcation
surface in the parameter space~\cite{VanCutsem1998, DobsonChiang1989}. Alternatively, the
Jacobian of only the network equations may become singular, leading to a singularity-induced bifurcation~\cite{Venkatasubramanian1995}. 
All of this treats demand as a
quasi-static parameter, for which the theory is mature.

\looseness=-1 Cyclic loads have been studied since the
1960s through linearized frequency-domain and time-simulation
techniques~\cite{VanNess1966, Rostamkolai1994}, which quantify
the forced oscillations but not the demand at which the periodic
steady state ceases to exist.
This letter derives the system loadability limit for this case by extending
the classical optimization argument of~\cite[Ch.~7]{VanCutsem1998} from
equilibria to fixed points of the period map. Three main results are obtained: First, at the limit,
the monodromy matrix has a Floquet multiplier at $+1$, hence the generic failure
is a cyclic fold. Second, this limit depends on the forcing frequency and, near
the electromechanical resonance, can lie far below the static margin. Third,
the basin of the stable orbit shrinks as the fold approaches and, as
a consequence, the operational margin is smaller than the existence
margin. A four-bus case study
demonstrates all three effects. 

%% file: sections/theory.tex
\vspace{-2mm}
\section{From Static to Periodic Loadability Limit}
\label{sec:theory}

\subsection{Static Argument}

We consider the differential-algebraic model
\begin{equation}
    \dot{x} = f(x, y, p), \qquad 0 = g(x, y, p),
    \label{eq:dae}
\end{equation}
with states $x \in \R^{n}$ and algebraic variables $y \in \R^{m}$ (bus
voltages), demand parameters $p$, and $g_y = \frac{\partial g}{\partial y}$ being nonsingular along the
trajectories of interest. Collect $u = (x, y)$ and
$\varphi(u, p) = (f, g)$, and let $\zeta(p)$ be a scalar demand measure
with $\nabla_p \zeta \neq 0$. The static loadability limit is obtained by
maximizing $\zeta(p)$ over all $(u, p)$ satisfying
$\varphi(u, p) = 0$~\cite[Ch.~7]{VanCutsem1998}. Its first-order
conditions are given by the stationarity of the Lagrangian
$L_0 = \zeta(p) + w^{\top} \varphi(u, p)$ with respect to $u$ and $p$,
\begin{equation}
    \varphi_u^{\top} w = 0, \qquad
    \nabla_p \zeta + \varphi_p^{\top} w = 0.
    \label{eq:static-kkt}
\end{equation}
Because $\nabla_p \zeta \neq 0$, the second condition in
\eqref{eq:static-kkt} excludes $w = 0$. From the first condition, it then follows that the square matrix
$\varphi_u$ has a nontrivial left null vector, hence,
$\det \varphi_u(u^{\ast}, p^{\ast}) = 0$ at the loadability limit, with
the asterisk marking quantities at this limit.
This singularity is the necessary condition for a saddle-node
bifurcation (SNB)~\cite{VanCutsem1998, DobsonChiang1989}, and the
demands admitting an equilibrium are bounded by the surface
$\Sigma_0 = \{p \,|\, \det \varphi_u = 0\}$ in the parameter space.
If an equipment limit becomes active first, it enters the optimization
as an additional constraint, and the limit-induced maximizer need not
be an
SNB~\cite[Ch.~7]{VanCutsem1998}.
\vspace{-0.45cm}
\subsection{Period Map and Cyclic-Fold Limit}

Let one component of the demand pulsate,
\begin{equation}
    p_d(t) = P_0 + P_f\, s(\theta), \qquad \dot{\theta} = 2\pi f_e,
    \label{eq:forcing}
\end{equation}
\looseness=-1 with $s$ a $2\pi$-periodic, zero-mean waveform, $|s| \le 1$, so that
$P_0$ is the mean demand and $P_f$ the peak deviation. Adjoining the
phase $\theta$ restores autonomy, but since $\dot\theta$ never
vanishes, the extended system has no equilibrium. Instead, the natural
steady state is a periodic orbit of period $T = 1/f_e$. Let
$\pmap(\chi; p)$ denote the flow of \eqref{eq:dae}, \eqref{eq:forcing}
over one period, starting from the physical states $\chi$ at a fixed
phase $\theta_0$, with the algebraic variables determined by $g = 0$.
An orbit is a fixed point of this period map,
\begin{equation}
    F(\chi, p) := \pmap(\chi; p) - \chi = 0,
    \label{eq:fixed-point}
\end{equation}
and fixing $\theta_0$ removes the phase-shift degeneracy and
the trivial $+1$ multiplier it generates. The derivative
of the map with respect to its starting point corresponds to the
monodromy matrix $M = \partial \pmap / \partial \chi$. Linearizing
\eqref{eq:dae} about the orbit, a small perturbation of the starting
state evolves as $\Delta\chi(t) = \Phi(t)\, \Delta\chi(0)$, where the
fundamental matrix $\Phi(t)$ solves the variational equation
\begin{equation}
    \dot{\Phi} = A(t)\, \Phi, \;\; \Phi(0) = I, \;\;
    A = \bigl( f_x - f_y g_y^{-1} g_x \bigr)\big|_{\text{orbit}},
    \label{eq:variational}
\end{equation}
with $A(t)$ the Jacobian of \eqref{eq:dae} along the orbit after the
algebraic perturbations are eliminated through $g = 0$. Integrating
\eqref{eq:variational} over one period gives $M = \Phi(T)$, which maps
a perturbation to its image one period later. If a limiter switches
during the cycle, such as the anti-windup field limit,
\eqref{eq:variational} is integrated piecewise and $\Phi$ is
multiplied at each switching instant by a saltation matrix, a
first-order correction for the jump in the vector
field~\cite{HiskensPai2000}. The eigenvalues of $M$ are the Floquet
multipliers $\lambda_1, \dots, \lambda_n$, and the orbit is
asymptotically stable when they lie inside the unit
circle~\cite{Kuznetsov2004}.

\looseness=-1 The periodic loadability limit is defined by the same optimization
problem with the
equilibrium constraint replaced by the orbit constraint, i.e., maximize
$\zeta(p)$ over all $(\chi, p)$ satisfying $F(\chi, p) = 0$.
Stationarity of $L = \zeta(p) + w^{\top} F(\chi, p)$ with respect to
$\chi$ and $p$ now requires
\begin{equation}
    F_\chi^{\top} w = (M - I)^{\top} w = 0, \qquad
    \nabla_p \zeta + F_p^{\top} w = 0,
    \label{eq:periodic-kkt}
\end{equation}
where $F_p = \partial \pmap / \partial p$. Because
$\nabla_p \zeta \neq 0$, the second condition in
\eqref{eq:periodic-kkt} again excludes $w = 0$ as a solution; from
the first condition it follows
that $M - I$ has a nontrivial left null vector, and therefore
\begin{equation}
    \det\bigl( M(\chi^{\ast}, p^{\ast}) - I \bigr) = \prod_{i=1}^{n} (\lambda_i-1)= 0
    \label{eq:fold}
\end{equation}
at the loadability limit of the forced system. Hence, a Floquet
multiplier lies at $+1$. A real multiplier crossing $+1$ corresponds precisely to a
cyclic fold, where a stable and an unstable
orbit collide and annihilate, and beyond which no orbit
exists~\cite{Kuznetsov2004}. As at the static limit, sensitivities
diverge: differentiating \eqref{eq:fixed-point} along the orbit branch
gives $(M - I)\, \partial \chi / \partial p + F_p = 0$, so the orbit
sensitivity $\partial \chi / \partial p = -(M - I)^{-1} F_p$
grows without bound as \eqref{eq:fold} is approached. Unlike the static
problem, active device limits require no reformulation here, because $M$
already embeds them through its saltation factors.

\looseness=-1 The static result is the zero-amplitude special case. For
$P_f = 0$ the orbit degenerates to the equilibrium, $A$ is constant,
$M = e^{A T}$, and a multiplier at $+1$ is a zero eigenvalue of $A$,
equivalent to singular $\varphi_u$ whenever $g_y$ is
nonsingular. For $P_f > 0$, however, $M$ is assembled from $A(t)$
around the entire cycle and through the period $T = 1/f_e$, such
that the
admissible demands are bounded by a surface $\Sigma$ in
$(P_0, P_f, f_e)$-space whose $P_f = 0$ slice is $\Sigma_0$.
Loadability under forcing is thus a function of frequency, which no
static computation can detect.

\subsection{Network Singularity and Reachability}

\looseness=-1 Condition \eqref{eq:fold} presumes that $g_y$ stays nonsingular along
the whole cycle. The loss of that property
at some instant of the cycle,
\begin{equation}
    \min_{t \in [0, T]} \smin\!\bigl( g_y \bigr)\big|_{\text{orbit}}
    \;\to\; 0,
    \label{eq:sib}
\end{equation}
with $\smin$ the smallest singular value, marks the orbit reaching
the singular surface of \eqref{eq:dae}, where the voltages cease to be
solvable. The corresponding static case of \eqref{eq:sib} is when the equilibria undergo singularity-induced
bifurcation~\cite{Venkatasubramanian1995}.
Whether \eqref{eq:fold} or \eqref{eq:sib} is met first depends on
damping, grid strength, and limiter status, as
Section~\ref{sec:case} shows.

\looseness=-1 Finally, existence does not imply reachability, because near the fold
the stable orbit coexists with an unstable companion whose stable
manifold bounds the basin of attraction, which shrinks to zero at the
fold. A
disturbance, or simply energizing the load from a power-flow state, can
therefore fail even when an admissible orbit provably exists.
Pseudo-arclength continuation~\cite{Keller1977} locates the existence
boundary independently of any basin; that boundary is what
\eqref{eq:fold} bounds.
\vspace{-0.35cm}

%% file: sections/casestudy.tex
\section{Numerical Validation}
\label{sec:case}

\looseness=-1 We use the four-bus system introduced
in~\cite{VanCutsem1998}, consisting of a stiff grid behind a reactance $X = 0.1$~p.u., one
500-MVA synchronous machine with the one-axis flux-decay model, inertia constant 3.5 s, and damping coefficient $D=0.4$~p.u. on machine base, with a first-order AVR (anti-windup
limits $E_{fd} \in [0, 5]$), and a load bus with a 13.5-p.u.\ active, 7.5-p.u.\ reactive exponential
load, plus a 6-p.u.\ shunt capacitor, and a data-center constant-power load with mean $P_0 =
    1.5$~p.u.\ and sinusoidal forcing~\eqref{eq:forcing}, all in per unit on
100~MVA. Orbits are determined by Newton shooting on
\eqref{eq:fixed-point}~\cite{AprilleTrick1972}, 
while \eqref{eq:fold} and \eqref{eq:sib} are both
monitored. Amplitudes are restricted to $P_f < P_0$, such that the load power stays
positive over the whole cycle.

\looseness=-1 The static comparator $P_{\mathrm{stat}}$ is obtained without periodic
forcing. The demand is raised quasi-statically from the operating
point, at 0.002~Hz, which is slow compared to all machine and AVR time
constants, so the trajectory tracks the equilibrium family of the
model with fixed control settings. The family ends at $P \approx 3.98$~p.u.,
where $E_{fd}$ reaches its ceiling and the limited model turns
unstable, a limit-induced instability~\cite{DobsonLu1992}, while
$\smin(g_y)$ stays bounded. The static margin above the mean demand is
therefore $\Delta P_{\mathrm{stat}} = 2.48$~p.u.

\begin{figure}[t!]
    \centering
    \includegraphics[width=\columnwidth]{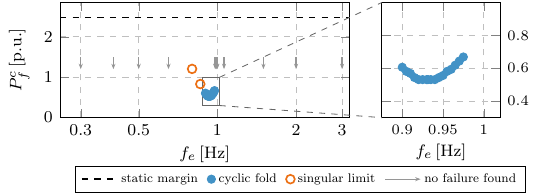}
    \vspace{-0.5cm}
    \caption{Critical forcing amplitude $P_f^{c}$ versus forcing
        frequency for the lightly damped, field-limited system
        ($D = 0.4$). Filled circles mark fold limits
        \eqref{eq:fold}, open circles singularity limits
        \eqref{eq:sib}, and downward arrows frequencies with no
        instability below the ceiling $P_f = P_0$. The dashed
        line is the static margin and the right panel enlarges the
        band around the swing frequency.}
    \label{fig:ampcrit}
    \vspace{-0.15cm}
\end{figure}

\looseness=-1 Fig.~\ref{fig:ampcrit} illustrates the result for
the cyclic loads and compares it to the static limit. For each forcing frequency
between 0.3 and 3~Hz, the amplitude $P_f$ was increased adaptively until
Newton shooting no longer converged. The limit was then bracketed to
0.02~p.u.\ by bisection, warm-started from the last
converged orbit, and the instability was read from whichever indicator
reached its threshold first, the real multiplier for
\eqref{eq:fold} or $\smin^2(g_y)$ for \eqref{eq:sib}. Frequencies
with no instability below the ceiling have no critical amplitude,
marked by downward arrows at $P_f = P_0$. A
limit below full modulation exists only in a band around the 0.93-Hz swing
frequency. Inside this band, the loadability limit shrinks to $P_f^{c} = 0.53$~p.u. (right panel), 4.7 times
below the static margin. The right edge is
abrupt, with a fold at 0.67~p.u.\ at 0.975~Hz but no instability at
0.98~Hz or above.
At 0.5~Hz, for instance, no instability occurs below the ceiling,
yet at 0.93~Hz the orbit is lost at a third of that
amplitude, and no static study distinguishes the two. Finally, the left panel shows instability through the network
singularity at 1.21 and 0.83~p.u.\ at 0.80 and 0.86~Hz, respectively.

\begin{figure}[t!]
    \centering
    \includegraphics[width=\columnwidth]{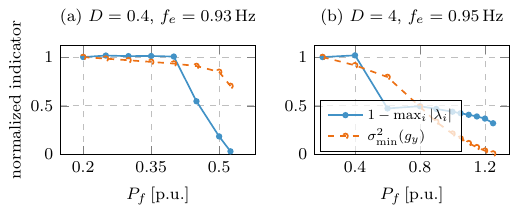}
    \vspace{-0.5cm}
    \caption{Amplitude sweep for the two indicators. (a)~Resonant,
        lightly damped case ($P_0 = 1.5$~p.u.), where the fold
        indicator reaches zero
        first. (b)~Strongly damped case ($P_0 = 3.0$~p.u.), where
        the singularity indicator reaches zero first.}
    \label{fig:race}
    \vspace{-0.1cm}
\end{figure}

\looseness=-1 Fig.~\ref{fig:race} shows both
indicators along the amplitude sweep at a nearly resonant period for two
damping levels. The values are normalized such that the first curve
to reach zero identifies the
mechanism of failure. With low damping (panel~(a)), the fold indicator $1 - \max_i
    |\lambda_i|$ collapses from a value of 1 to zero within a 0.1-p.u.\ amplitude
window, while the
dominant multiplier, as defined in \eqref{eq:fold}, stays real and
positive. As the
network is still far from singularity, the orbit dies as a rotor-angle
instability. In the strongly damped case (panel~(b)), the fold
indicator decreases once the field limit becomes active during part
of the cycle, but it stays far
from zero. Instead, $\smin^2(g_y)$ decreases linearly to zero at $P_f
    \approx 1.27$~p.u., indicating that \eqref{eq:sib} is the limiting
factor. The two limits call for
different countermeasures, damping in one case and reactive power support in the
other, and a screening tool that tests only one certifies the wrong margin when
the other binds first.

\begin{figure}[t!]
    \centering
    \includegraphics{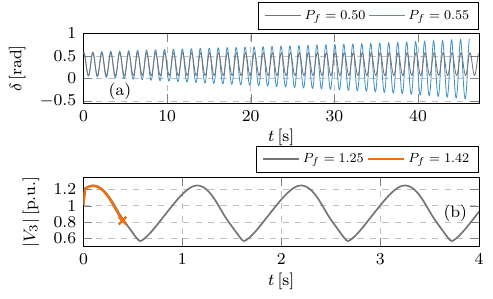}
    \vspace{-0.5cm}
    \caption{Instability signatures. (a)~Rotor angle at resonance for
        $P_f = 0.50$ (stable orbit) and $P_f = 0.55 > P_f^{c}$,
        diverging after 43
        periods. (b)~Load-bus voltage for the strongly damped case
        ($P_0 = 3.0$~p.u.) at $P_f = 1.25$ (orbit) and $P_f = 1.42$,
        losing solvability within the first cycle (cross).}
    \label{fig:traces}
    \vspace{-0.15cm}
\end{figure}

\looseness=-1 Fig.~\ref{fig:traces} confirms the two mechanisms in the time
domain, showing simulated trajectories that bracket the limit; both
are integrated from the last stable
orbit and thus start from the same state. In Fig.~(a),
the amplitude steps across the resonant fold from 0.50 to 0.55~p.u. Below the
fold, the response is a bounded hybrid orbit whose field voltage $E_{fd} \in [0, 5]$ saturates at both
limits during every cycle. Above the fold, no orbit exists, yet the rotor swings
grow so slowly that synchronism is lost only after 43 forcing
periods. In Fig.~\ref{fig:traces}(b), the amplitude steps across the singularity limit of the
strongly damped case when increasing $P_f$ from 1.25 to 1.42~p.u.
For $P_f = 1.42$~p.u.\ the network equations become singular within
the first cycle, in 0.39~s. The fold in Fig.~\ref{fig:traces}(a) thus gives tens of seconds of apparently
periodic operation on an orbit that no longer exists, while the singularity
leaves almost no time to react.

\looseness=-1 The last experiment compares two amplitudes: the
largest that operation can actually reach, and the largest at which
the orbit still exists.
The reachable margin is found by energizing the load from a power-flow
start and bisecting the largest amplitude that settles onto the orbit
within 40 periods. The existence boundary is found by
pseudo-arclength continuation. At the sharp
resonant fold, with $D = 0.4$ and $f_e = 0.93$~Hz, the two nearly
coincide, namely 0.520 versus 0.527~p.u. As a counterexample, with
$D = 1.2$ at $f_e = 0.952$~Hz, cold starts fail beyond $P_f = 0.90$~p.u. while the continued
branch persists past 1.13~p.u., indicating that at least a fifth of the existence margin is
operationally unreachable.

%% file: sections/conclusion.tex
\vspace{-2mm}
\section{Conclusion}

\looseness=-1 In this letter, we derived the loadability limit of a power system
whose demand contains a periodic component. The classical optimization
argument carries over from equilibria to periodic orbits, with the
role of the equilibrium Jacobian taken by the monodromy matrix.
Consequently, the generic limit is a cyclic fold, at which a Floquet
multiplier reaches unity. The static saddle-node is
recovered at zero periodic amplitude. Moreover, a second limit arises where the
orbit meets the surface on which the network equations lose
solvability. Both limits depend
on the forcing frequency. Near resonance it can lie significantly
below the static margin, and part of the existence margin is
unreachable in operation. For data-center-scale periodic loads,
interconnection screening should therefore test the cyclic-fold and
the network-singularity conditions over the credible forcing-frequency
band rather than a single static nose.
\vspace{-0.35cm}